\documentclass[final]{aipproc}

\layoutstyle{6x9}

\usepackage{amsmath, amsfonts, amssymb}
\usepackage{graphicx}
\usepackage{soul}
\usepackage{array}
\usepackage[section]{placeins}
\usepackage{wrapfig}

\def\lag{\langle}
\def\rag{\rangle}
\def\c{\chi}

\begin{document}

\title{Accommodate chiral symmetry breaking and linear confinement in
a dynamical holographic QCD model}


\author{Danning Li}{
address={Institute of High Energy Physics, Chinese Academy
of Sciences, Beijing, China }
}

\author{Mei Huang}{
address={Institute of High Energy Physics, Chinese Academy
of Sciences, Beijing, China }
,altaddress={Theoretical Physics Center for Science Facilities, Chinese
Academy of Sciences, Beijing, China}
}

\author{Qi-Shu Yan}{
address={College of Physical Sciences, Graduate
University of Chinese Academy of Sciences, Beijing, China}
}

\begin{abstract}
We construct a self-consistent holographic QCD model which can realize two most important phenomena of QCD, i.e. chiral symmetry breaking and confinement.
The model is formulated in the framework of graviton-dilaton-scalar system,
where the dilaton field is of dimension-2 which might be dual to the
dimension-2 gluon condensate and can lead to the linear confinement, while
the scalar field corresponds to the quark anti-quark condensate and can explain
the property of chiral dynamics. Within this framework, both Regge spectra of
hadrons and the linear potential between quarks can be accommodated. It is also found that the negative dilaton background can be safely excluded in this framework.
\end{abstract}

\maketitle


\section{Introduction}

It is well-known that the QCD vacuum is characterized by spontaneous
chiral symmetry breaking and color charge confinement. The spontaneous
chiral symmetry breaking is well understood by the dimension-3 quark condensate
$\lag\bar{q}q\rag$ \cite{NJL} in the vacuum, in spite of that, the understanding
to confinement remains a challenge. Confinement can be reflected by the Regge
trajectories of hadrons \cite{Regge}, which suggests that the color charge can
form the string-like structure inside hadrons. It can also be shown by the
linear potential between two quarks (either light or heavy) at large distances,
{\it i.e.} $V_{\bar{Q}Q}(R)=\sigma_s R$ with $\sigma_s$ the string tension.

The discovery of the anti-de Sitter/conformal field theory (AdS/CFT)
correspondence and the conjecture of the gravity/gauge duality
\cite{dual} provides a revolutionary method to tackle the problem of
strongly coupled gauge theories. Many efforts have been invested in
searching for such a realistic description by using the "bottom-up"
approach, e.g. see Ref. \cite{review} for reviews.
The Regge trajectories and linear quark potential are like two sides
of a coin, both describe the properties of linear confinement.
A successful holographic QCD model
should describe both the Regge trajectories of hadron spectra and linear quark potential. Nonetheless,
the models on market cannot accommodate both.
Currently a working framework used to describe the Regge
trajectories of hadron spectra is the soft-wall AdS$_5$ model or
Karch-Katz-Son-Stephanov (KKSS) model \cite{Karch:2006pv} and
its extended version \cite{Gherghetta-Kapusta-Kelley,YLWu,modified-dc},
where a quadratic dilaton background is introduced in the 5D action.
However, with AdS$_5$ metric, only Coulomb potential between the two
quarks can be produced \cite{Maldacena:1998im}. The working holographic
QCD model to realize the linear quark potential was proposed in
Ref.\cite{Andreev:2006ct}, where a positive quadratic correction in
the deformed warp factor of ${\rm AdS}_5$ geometry was introduced.
(The linear heavy quark potential can also be obtained by introducing other deformed warp
factors as shown in Refs. \cite{Pirner:2009gr,He:2010ye}.) The positive
quadratic correction in the deformed warp factor in some sense behaves as a
negative dilaton background in the 5D action, which motivates the
proposal of the negative dilaton soft-wall model \cite{Zuo:2009dz,deTeramond:2009xk}.
More discussions on the sign of the dilaton correction can be found in \cite{Schmidt-pn-dilaton,KKSS-2}.

It is interesting to explore how to generate both the linear Regge
behavior of hadron spectra and linear quark potential in a
self-consistent model. In this work, a dynamical holographic QCD model is
proposed and formulated in the graviton-dilaton-scalar coupled system \cite{Li:2012ay}.

\section{Graviton-dilaton-scalar coupled system}

We first briefly describe the graviton-dilaton coupled system, where
the dilaton background is expected to be dual to the effective degree of
freedom in the pure gluon system. In general, a dilaton background will
deform the warp factor of the metric structure \cite{Csaki:2006ji,GKN,Li:2011hp}.
The 5D action of the graviton-dilaton system is defined as
\begin{equation}\label{action-GD}
S_{GD}= \frac{1}{16\pi G_5}\int
 d^5x\sqrt{g_s}e^{-2\phi}\big(R+4\partial_m\phi\partial^m\phi-V_{\phi}\big).
\end{equation}
Where $G_5$ is the 5D Newton constant, $g_s$, $\phi$ and $V_\phi$ are the 5D
metric, the dilaton field and dilaton potential in the string frame, respectively.
Under the quadratic dilaton background
$\phi=\mu^2 z^2$,
the analytic solution of the dilaton potential in the Einstein frame $V^E_\Phi
=e^{4\phi/3}V_{\phi}$ with $\phi=\sqrt{\frac{3}{8}}\Phi$ takes the form of
\begin{equation}
V_{\Phi}^E=-12~\frac{ _0F_1(1/4;\frac{\Phi^2}{24})^2}{L^2}
+2~ \frac{ _0F_1(5/4;\frac{\Phi^2}{24})^2\Phi^2}{L^2},
\end{equation}
here $L$ the radius of AdS$_5$ and $_0F_1(a;z)$ the hypergeometric function.
In the ultraviolet limit,
$V^E_{\Phi}\overset{\Phi\rightarrow0}{\longrightarrow}-\frac{12}{L^2}+\frac{1}{2}M^2_{\Phi}\Phi^2$
with the 5D mass for the dilaton field $M^2_{\Phi}L^2=-4$. From the
AdS/CFT dictionary $\Delta(\Delta-4)=M^2_{\Phi}L^2$, one can derive
its dimension $\Delta=2$. The most likely dimension-2 operator
candidate in QCD is $A_{\mu}^2$. It has been pointed out in
Ref.\cite{GC-D2} that the dimension-2 gluon condensate
plays essential role for the linear confinement. This  $A_{\mu}^2$ can be put into
a gauge invariant form \cite{Kondo}, which might be related to topological
defects in QCD vacuum \cite{Gubarev:2000nz}. The quadratic
dilaton field might be dual to the dimension-2 gluon condensate,
and the graviton-dilaton system describes the pure gluodynamics.

We now add the probe of flavor dynamics on the pure gluodynamic background,
and extend the graviton-dilaton system to the framework of graviton-dilaton-scalar
coupling system, where the scalar field captures chiral dynamics.
The graviton-dilaton-scalar system can be described by the following 5D action:
\begin{equation}
S=S_{GD}+S_{M},
\label{totalaction}
\end{equation}
with $S_{GD}$ given in Eq.(\ref{action-GD}) and $S_{M}$ the KKSS
action for mesons as in \cite{Karch:2006pv} taking the form of
\begin{equation}\label{action1st}
S_{M} = -\frac{N_f}{N_c} \int d^5x
 \sqrt{g_s} e^{-\phi} Tr\Big(|DX|^2+V_{X}  +\frac{1}{4g_5^2}(F_L^2+F_R^2)\Big).
\end{equation}
Where $X$ and $V_{X}$ are the scalar field and its corresponding potential.
$g_5^2$ is fixed as $12\pi^2N_f/N_c^2$ \cite{Karch:2006pv} and we take
$N_f=2,N_c=3$ in this paper.

We assume the vacuum background is induced by the dilaton field of dimension-2 gluon
condensate and the scalar field of the quark antiquark condensate
$<X>=\frac{\c(z)}{2}$ \cite{Karch:2006pv}, then the vacuum background part of the
action Eq.(\ref{totalaction}) takes the following form
\begin{eqnarray}\label{actionbgrs}
S_{vac}&=&\frac{1}{16\pi G_5}\int d^5x\sqrt{g_s}\big\{
e^{-2\phi}\big(R_s+4\partial_m\phi\partial^m\phi-V_{\phi}\big)\nonumber\\
&-&\lambda e^{-\phi}\big(
\frac{1}{2}\partial_m\c\partial^m\c+V_{\c} \big)\big\},
\end{eqnarray}
with $\lambda=\frac{16\pi G_5 N_f}{N_c L^3}$ and the metric in the
string frame
\begin{equation}
dS_s^2= B_s^2(-dt^2+d\overset{\rightarrow}{x}^2+dz^2), ~~~ B_s^2\equiv e^{2A_s}\equiv L^2 b_s^2.
\end{equation}

It is easy to derive the following three coupled field equations:
\begin{eqnarray}
 -A_s^{''}+A_s^{'2}+\frac{2}{3}\phi^{''}-
 \frac{4}{3}A_s^{'}\phi^{'}-\frac{\lambda}{6}e^{\phi}\c^{'2}&=&0, \label{eom-aphikai} \\
\phi^{''}+(3A_s^{'}-2\phi^{'})\phi^{'}-\frac{3\lambda}{16}e^{\phi}\chi^{'2} & & \nonumber \\
 -\frac{3}{8}e^{2A_s-\frac{4}{3}\phi}\partial_{\phi}\left(e^{4/3\phi}V_{\phi}
 +\lambda e^{7/3\phi}V_{\c}\right)&=&0, \label{eom-phi} \\
 \c^{''}+(3A_s^{'}-\phi^{'})\c^{'}-e^{2A_s}\partial_{\c}V_{\c}&=&0.
 \label{eom-kai}
\end{eqnarray}
If we know the dynamical information of the dilaton field $\phi$ and the scalar field $\chi$,
then the metric $A_s$, the dilaton potential $V_{\phi}$ and the scalar potential $V_{\c}$
should be self-consistently solved from the three coupled equations given in Eqs.(\ref{eom-aphikai},\ref{eom-phi},\ref{eom-kai}).
It is noticed that the graviton-dilaton-scalar system is different from the 
graviton-dilaton-tachyon system \cite{GDT}, where the metric remains as AdS$_5$.

\subsection{Constraints from chiral symmetry breaking and linear confinement}

As proposed in the KKSS model, at ultraviolet(UV), the scalar field takes the
following asymptotic form,
\begin{eqnarray}
\chi(z) \stackrel{z \rightarrow 0}{\longrightarrow} m_q \zeta z+\frac{\sigma}{\zeta} z^3,
\label{chi-IR}
\end{eqnarray}
where $m_q$ is the current quark mass, and $\sigma$ is the quark
antiquark condensate, and $\zeta$ is a normalization constant and is
fixed as $\zeta^2=\frac{N_c^2}{4\pi^2N_f}$. In this paper, we would
fix $m_q=5 {\rm MeV},\sigma=(228 {\rm MeV})^3$ .

The linear behavior of quark-antiquark static potential
in the heavy quark mass limit $m_Q\rightarrow \infty$ can describe
the permanent confinement property of QCD. Following
Ref. \cite{Maldacena:1998im}, one can solve
the renormalized free energy of the $\bar{q}q$ system
under the general metric background $A_s$.
One can find that at the point $z_c$ when $b_s^{'}(z_c)\rightarrow 0$,
\begin{eqnarray}
\frac{V_{\bar{q}q}(z_0)}{R_{\bar{q}q}(z_0)}\overset{z_0\rightarrow z_c}{\longrightarrow} \frac{g_p}{2\pi} b_s^2(z_c). 
\label{stringtension}
\end{eqnarray}
Here $g_p=\frac{L^2}{\alpha}$ and $\alpha$ the 5D effective string tension.
Therefore, the necessary condition for the linear part of the
$q-\bar{q}$ potential is that there exists one point $z_c$ or one region,
where $b_s^{'}(z)\rightarrow 0,z\rightarrow z_c$ while $b_s(z)$ keeps
finite. For simplicity, we can take the following constraint on the
metric structure at IR:
$A_s^{'}(z) \stackrel{z \rightarrow \infty}{\longrightarrow} 0,
A_s(z) \stackrel{z \rightarrow \infty}{\longrightarrow} {\rm Const}$.
Under these conditions,
the equation of motion Eq.(\ref{eom-aphikai}) in the IR takes a simple form:
$\frac{2}{3}\phi^{''}-\frac{\lambda}{6}e^{\phi}\c^{'2}=0$,
which provides a relation between the chiral condensate and
dimension-2 gluon condensate at IR. The asymptotic form of
$\c(z)$ at IR can be solved as:
\begin{equation}
\chi(z)\stackrel{z \rightarrow \infty}{\longrightarrow} \sqrt{8/\lambda}\mu e^{-\phi/2}.
\label{chi-UV}
\end{equation}

To match the asymptotic forms both at UV and IR in Eqs.(\ref{chi-IR}) and (\ref{chi-UV}),
$\c$ can be parameterized as
\begin{eqnarray}\label{chiz}
\c^{'}(z)=\sqrt{8/\lambda}\mu e^{-\phi/2}(1+c_1 e^{- \phi}+c_2
e^{-2\phi})
\end{eqnarray}
with
$c_1=-2+\frac{5\sqrt{2\lambda}m_q\zeta}{8\mu}+\frac{3\sqrt{2\lambda}\sigma}{4\zeta
\mu^3}$ and $c_2=1-\frac{3\sqrt{2\lambda}m_q\zeta}{8\mu}-\frac{3\sqrt{2\lambda}\sigma}{4\zeta
\mu^3}$.

\subsection{Regge trajectories of mesons}

Under the positive quadratic dilaton background and the scalar profile Eq.(\ref{chiz}), the metric 
structure $A_s(z)$ or $b_s(z)$ can be solved through Eq.(\ref{eom-aphikai}).
Considering the meson fluctuations under the vacuum background described by 
$A_s, \phi, \c$ as in \cite{Karch:2006pv}, we can split the fields into background part
and perturbation part. For $X$ we would have a scalar perturbation
$s$ and pseudo-scalar perturbation
$\pi$, i.e., $X=(\frac{\c}{2}+s)e^{i 2\pi^a t^a}$. For the vector
field $V_\mu$ and axial vector field $A_\mu$ part, due to their
vanishing vacuum expectation value, we would use the same notation to denote
the perturbation fields. The equations of motion for perturbation fields
$s,\pi,V_\mu,A_\mu$ can be easily derived. For example, the schrodinger
like equation for vector is given below:
\begin{eqnarray}
& &-v_n^{''}+V_v(z)v_n=m_{n,v}^2v_n, \\
\label{scalar-sn}
& & V_v(z)=\frac{A_s^{''}-\phi^{''}}{2}+\frac{(A_s^{'}-\phi^{'})^2}{4} \label{scalar-vsz} .
\end{eqnarray}
It is noticed that at IR, $V_v(z)=-\mu^2 + \mu^4z^2$,
therefore the solution of Eq.(\ref{scalar-sn}) for high excitations is
$M_n^2=4\mu^2 n$.

\begin{figure}
\includegraphics[height=.3\textheight]{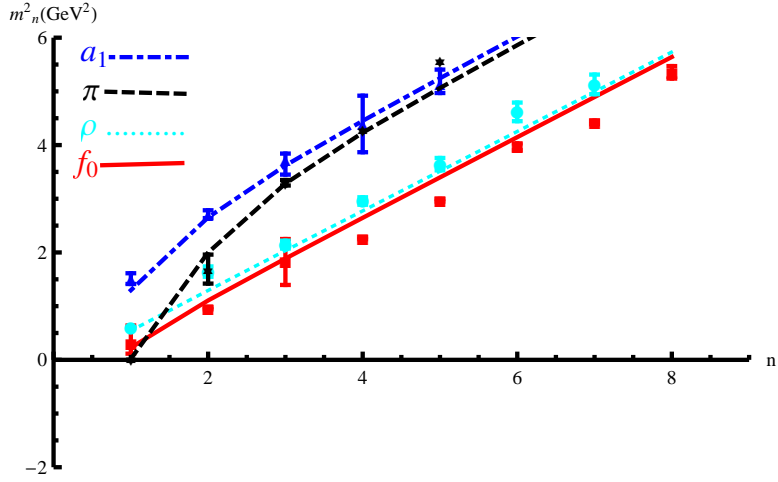}
\caption{A plot of experimental(dot) and model predicted (line) mass
square spectra for the scalar and pseudoscalar mesons $f_0,\pi$ and
vector and axial-vector mesons $\rho, a_1$.} \label{pif0rhoa1-mass}
\end{figure}

By fixing the 5D Newton constant $G_5=\frac{3 L^3}{4}$, one can produce the proper
splitting between the vector and axial vector Regge trajectories.
The produced spectra of scalar $f_0$, pseudoscalar $\pi$, vector $\rho$ and axialvector $a_1$
are shown in Fig.\ref{pif0rhoa1-mass} compared with experimental data \cite{pdg}. The experimental 
data for $f_0$ are chosen as in Ref.\cite{Gherghetta-Kapusta-Kelley}. With only 4 parameters, all 
produced meson spectra in the graviton-dilaton-scalar system agree well with experimental data. 
We would emphasize that our model can incorporate both chiral symmetry breaking and confinement 
properties in the hadron spectra, and the slope of Regge trajectories is $4\mu^2$ with $\mu=0.43 {\rm GeV}$. 
In a similar dynamical holographic QCD model in Ref. \cite{dePaula:2008fp}, the linear Regge behavior is 
realized but the chiral symmetry breaking mechanism is missing.

\subsection{The string tension of the linear quark potential.}

\begin{figure}
\includegraphics[height=.3\textheight]{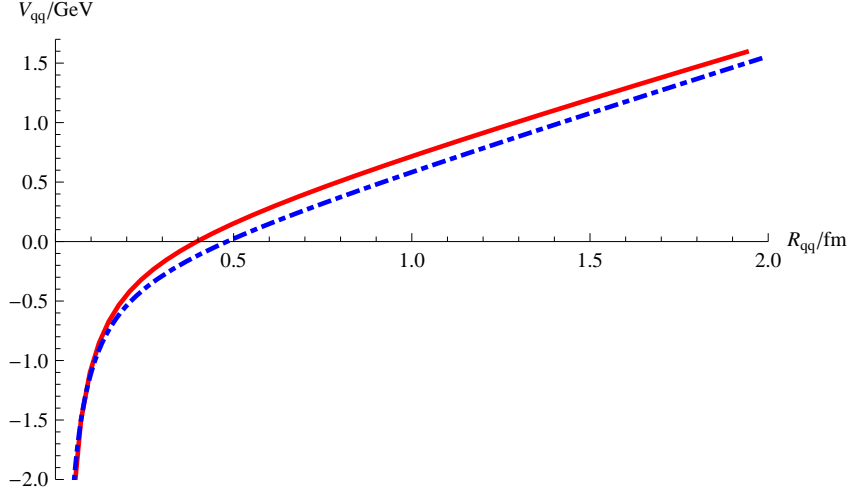}
\caption{$V_{\bar{q}q}$ as a function of $R_{\bar{q}q}$ from our
model (solid line) with $g_p=2.3$ and $\mu=0.43{\rm GeV}$ compared
with the Cornell potential (dot-dashed line). } \label{Vqq}
\end{figure}

Under the metric background $A_s(z)$ solved from Eq.(\ref{eom-aphikai}) with the positive 
quadratic dilaton background and the scalar profile Eq.(\ref{chiz}), the quark potential
can be solved numerically. In the UV limit, one can derive the 
Coulomb potential $V_{\bar{q}q}=-\frac{0.23 g_p}{R_{\bar{q}q}}$ as given in Ref.\cite{Maldacena:1998im}. 
In the IR limit, we can get the linear potential $V_{\bar{q}q}=\frac{g_p}{2\pi} b_s^2(z_c) R_{\bar{q}q}$. 
From the solutions in Eq.(\ref{eom-aphikai}), we have $b_s^2 \approx 4 \mu^2$, which indicates that 
the string tension of the linear quark potential $\sigma_s \sim 4 \mu^2$.
The numerical result for the quark potential $V_{\bar{q}q}$ as a function
of quark anti-quark distance $R_{qq}$ is shown by the solid line in
Fig.\ref{Vqq}. The two parameters  are fixed as
$g_p=2.3$ and $\mu=0.43 {\rm GeV}$. The result agrees with the
Cornell potential (dot-dashed line) \cite{Cornell}
$V^c(R)=-\frac{\kappa}{R}+\sigma_{str}R+V_0$  with $\kappa\approx
0.48$, $\sigma_{str}\approx 0.183 {\rm GeV}^{2}$ and $V_0=-0.25 {\rm
GeV}$.

\subsection{The sign of the dilaton field}

The last but not the least, we address the issue of the sign of the dilaton background.
If we choose a negative dilaton background $\phi=-\mu^2 z^2$ as in 
Ref.\cite{Zuo:2009dz,deTeramond:2009xk}, in the IR limit,
the last term in Eq.(\ref{eom-aphikai}) decreases exponentially to zero, and 
one can get the asymptotic solution of $A_s(z) \overset{z\rightarrow\infty}{\longrightarrow} -\frac{4}{3}\mu^2z^2+1/2\log(z)$ and
$b_s(z) \sim \sqrt{z} e^{-\frac{4}{3}\mu^2 z^2}\overset{z\rightarrow\infty}{\longrightarrow}0$.
From Eq.(\ref{stringtension}), it's not possible to produce the linear potential with 
a negative dilaton background. Therefore, one can safely exclude the negative dilaton 
background in the graviton-dilaton-scalar system.

\section{Conclusion}

In summary, we propose a dynamical holographic QCD model, which takes into account  the 
back-reaction of flavor dynamics on the pure gluodynamic background. To our knowledge, this 
is the first dynamical holographic QCD model which can
produce both the linear Regge trajectories of hadron
spectra and quark anti-quark linear potential. It is observed that
both the slope of the Regge trajectories and the string tension of
the linear quark anti-quark potential are proportional to the dimension-2
gluon condensate. This result indicates that the linear confinement is
dynamically induced by the dimension-2 gluon condensate. The holographic QCD model 
offers us a new viewpoint on the relation between the chiral symmetry breaking and
confinement. It is found that the balance between the chiral
condensate and dimension-2 gluon condensate is essential to produce
the correct Regge behavior of hadron spectra. As a byproduct, it is
found that the negative dilaton background can be safely excluded in the
framework of graviton-dilaton-scalar system.

\section{Acknowledgements}

We thank valuable discussions with P. Colangelo,  M. Chernodub, M. Frasca, K. Kondo,
M. Ruggieri and A. Vairo during the QCD@Work.
This work is supported by the NSFC under Grant
Nos. 11175251 and 11275213, DFG and NSFC (CRC 110),
CAS key project KJCX2-EW-N01, K.C.Wong Education Foundation, and
Youth Innovation Promotion Association of CAS.

\bibliographystyle{aipproc}   

\end{document}